\documentstyle[12pt,aasms4]{article}

\lefthead{Kenji Bekki and Masashi Chiba}
\righthead{Formation of the Galactic stellar halo}

\begin{document}
\title{Formation of the Galactic stellar halo: Origin of
the metallicity-eccentricity relation}

\author{Kenji Bekki} 
\affil{Division of Theoretical Astrophysics,
National Astronomical Observatory, Mitaka, Tokyo, 181-8588, Japan} 

\and

\author{Masashi Chiba}
\affil{
Division of Astrometry and Celestial Mechanics,
National Astronomical Observatory, Mitaka, Tokyo, 181-8588, Japan}

\begin{abstract}
Motivated by the recently improved knowledge on the kinematic and chemical
properties of the Galactic metal-poor stars, we present the numerical
simulation for the formation of the Galactic stellar halo to interpret the
observational results. As a model for the Galaxy contraction, we adopt the
currently standard theory of galaxy formation based on the hierarchical
assembly of the cold dark matter fluctuations. We find, for the simulated
stars with [Fe/H]$\le-1.0$, that there is no strong correlation between metal
abundances and orbital eccentricities, in good agreement with the observations.
Moreover, the observed fraction of the low eccentricity stars is reproduced
correctly for [Fe/H]$\le-1.6$ and approximately for the intermediate abundance
range of $-1.6<$[Fe/H]$\le-1.0$. We show that this successful reproduction of
the kinematics of the Galactic halo is a natural consequence of the
hierarchical evolution of the subgalactic clumps seeded from the cold dark
matter density fluctuations.
\end{abstract}


\section{Introduction}

Structure and dynamics of the metal-deficient halo component in the Galaxy
provide valuable information on the early evolution of the Galaxy (e.g.,
Freeman 1987; Majewski 1993). Accordingly, the origin of the Galactic stellar
halo has been extensively discussed by many authors since Eggen, Lynden-Bell,
\& Sandage (1962, hereafter referred to as ELS) reported a strong correlation
between metal abundances and space motions of the high-velocity stars in the
solar neighborhood. ELS argued that the contraction of the Galaxy must have
been monolithic and rapid within a free-fall time ($\sim$ $10^{8}$ yr).
Several authors however pointed out that the collapse time scale estimated
by ELS is greatly affected by their selection criterion against the halo stars
having high angular momentum (e.g., Yoshii \& Saio 1979; Norris, Bessell,
\& Pickles 1985; Chiba \& Yoshii 1998, hereafter CY).
Alternatively, Searle \& Zinn (1978, hereafter SZ) proposed that the Galactic
halo was formed slowly ($\sim$ $10^{9}$ yr) by chaotic merging/accretion of
several subgalactic fragments. It is yet unsettled whether either
ELS's monolithic or SZ's merging picture (or both, e.g., Norris 1994; Freeman
1996; Carney et al. 1996) is more plausible and realistic for describing
the early evolution the Galaxy.

Kinematic aspects of the Galactic metal-poor stars have been greatly improved
by the recently completed $Hipparcos$ Catalogue (ESA 1997) and various
ground-based catalogs (e.g., Platais et al. 1998; Urban et al. 1998) which
provide unprecedentedly accurate proper motion data for a wealth of metal-poor
stars (Beers et al. 2000).
Using the non-kinematically selected sample of stars having available
proper motions, CY and Chiba \& Beers (2000, hereafter CB) revisited the
relation between metal abundances and orbital eccentricities
of the halo stars and found no evidence for a strong correlation between these
quantities, in contrast to ELS's finding.
CB also showed a clear evidence for internal structure of the halo: the outer
halo shows no systematic rotation and nearly spherical density distribution,
whereas the inner halo has a prograde rotation and a highly flattened density
distribution. Also, Helmi et al. (1999) discovered a statistically significant
clumpiness of the nearby halo stars in the angular momentum space, and argued
that about 10 \% of the halo come from a single small galaxy that was disrupted
during or soon after the Galaxy formation.
These new findings not only provide constraints on the formation of the
Galactic stellar halo but also improve our understanding of how disk galaxies
like our own form.

In this Letter, we address the question whether the above kinematic and
chemical properties of the Galactic halo are understandable in the context of
the currently favored theory of galaxy formation based on the hierarchical
assembly of cold dark matter (CDM) halos (White \& Rees 1978).
Most of the CDM-based numerical models on disk galaxy formation have focused
on only the fundamental properties of a disk, such as an exponential density
profile (Katz 1992) and Tully-Fisher relation (Steinmetz \& Navarro 1999).
The spatial structure of the stellar halo has been examined by Steinmetz \&
M\"uller (1995), but the detailed internal kinematics of the halo stars in the
simulated model remained unknown.
Here we explore a numerical simulation for the formation of
the Galactic halo, to investigate whether the CDM model can successfully explain
the kinematic and chemical properties of the Galactic halo.
We particularly focus on the evolution of hierarchically clustered subgalactic
clumps seeded from the CDM density fluctuations and investigate their
evolutionary effects on the orbital distribution of the stars in conjunction
with the metal enrichment. More extensive analyses and results of the
numerical simulations will be presented elsewhere (Bekki \& Chiba 2000).

\section{Model}

The numerical method and technique for solving galactic chemodynamical
evolution and models for describing star formation and dissipative gas dynamics
are presented in Bekki \& Shioya (1998), and here we briefly describe the
initial conditions of protogalactic clouds, star formation law, and chemical
evolution model. The way to set up initial conditions for numerical simulations
of forming disk galaxy within a hierarchical clustering scenario is essentially
the same as that adopted by Katz \& Gunn (1991) and Steinmetz \& M\"uller
(1995). 
We here use the COSMICS (Cosmological Initial Conditions and
Microwave Anisotropy Codes), which is a package
of fortran programs for generating Gaussian random initial
conditions for nonlinear structure formation simulations
(Bertschinger 1995; Ma \& Bertschinger 1995).
We consider an isolated homogeneous, rigidly rotating sphere, on which
small-scale fluctuations according to a CDM power spectrum are superimposed.
The mass ratio of dark matter to baryons is 9.0
and both components have the same initial distribution. 
The initial total mass, radius, and spin parameter ($\lambda$) of the sphere
in the present model are $6.0\times10^{11}\rm M_{\odot}$, 45 kpc, and
0.08, respectively. For cosmological parameters of $\Omega =1.0$, $q_0=0.5$,
and $H_{0}=50$ km $\rm s^{-1}$ ${\rm Mpc}^{-1}$ adopted in the present study,
the initial overdensity ${\delta}_{i}$ of the sphere in the fiducial model is
estimated to be 0.29 corresponding to a 2.5 $\sigma$ overdensity
of a biased CDM spectrum on a mass-scale of 6.0 $\times 10^{11} \rm M_{\odot}$ 
with a biasing parameter $b = 2$.
We start the simulation at the redshift $z = 25$ and follow it 
till $z=0$  corresponding to the age of the universe equal to  13 Gyr.
Initial conditions similar to those adopted in the present study
are demonstrated to be plausible and realistic for the formation
of the Galaxy (e.g., Steinmetz \& M\"uller 1995).

Star formation is modeled by converting the collisional gas particles into
collisionless new stellar particles according to the Schmidt law (Schmidt 1959)
with the exponent of 2 and the coefficients in the law are taken from
the work of Bekki (1998).
The collisional and dissipative nature of the interstellar medium are modeled
according to the sticky particle method (Schwarz 1981)
with the cloud radius ($r_{cl}$) of 450 pc and we consider multiple collisions
among clouds (see Bekki \& Shioya 1998 for details). The
total particle number used for modeling the initial sphere is 14147 both for
dark matter and for baryons (gas and new stars), 
which means that the mass of each particle is
3.8 $\times$ $10^7$ $ \rm M_{\odot}$ 
for dark matter and 4.2 $\times$ $10^6$ $ \rm M_{\odot}$ for baryons.
We assume the local mixing of metals, in such a way that those produced by
a new star are instantaneously assigned to
the gas particles located within $2r_{cl}$ from the star. 
The fraction of gas returned to interstellar medium in each
stellar particle and the chemical yield are 0.3 and 0.02, respectively.
All the calculations related to the above chemodynamical evolution
have been carried out on the GRAPE
board (Sugimoto et a. 1990)  at Astronomical Institute of Tohoku University.
The parameter of gravitational softening is set to be fixed at 0.053 in our
units (2.4 kpc). The time integration of
the equation of motion is performed by using the 2nd-order leap-flog method.

Using the above model, we derive the distribution of metal abundances ([Fe/H])
and orbital eccentricities ($e$) of the stars with [Fe/H]$\le-0.6$ at the
epoch $z = 0$. Here $e$ for each stellar particle is defined as:
\begin{equation}
  e= \frac{r_{apo}-r_{peri}}{r_{apo}+r_{peri}} \;
\end{equation}
where $r_{apo}$ and $r_{peri}$ are apo-galactic and peri-galactic distances
from the center of the simulated Galaxy, respectively.
For estimating $e$, we first select the stellar particles with [Fe/H]$\le-0.6$
found at $z=0$. Then we calculate the time evolution of their orbits
under the gravitational potential of the simulated disk Galaxy achieved at
$z = 0$ for ten dynamical time scale ($\sim$ 1.8 Gyr), and then estimate $e$.
In order to avoid the contamination of metal-poor 
bulge stars in this estimation,
we select only particles with their apo-galactic distances ranging from 
8.5 to 17.5 kpc.
In the followings, we use the symbol [Fe/H] as the total metal abundance
instead of the symbol Z to avoid confusion with redshift, although
the current model does not consider the evolution of each element
separately.

\placefigure{fig-1}
\placefigure{fig-2}

\section{Result}
Figure 1 shows the dynamical evolution of the star-forming gas sphere of the
Galaxy in the present CDM model. The largest density maxima within the initial
gas sphere become first non-linear ($z \sim 10$) and then collapse
to form subgalactic clumps consisting of gas and new stars. Firstly born stars
with old ages ($>$ 10 Gyr) and low metallicities ([Fe/H]$<-3$)
are located within these clumps till they are disrupted by later mutual
merging. The initial gas sphere reaches the turn-around point at $z \sim 3.25$
and then begins to collapse. Subgalactic clumps developed from local
small-scale density perturbations within the gas sphere fall onto the inner
region of the proto-Galaxy owing to dynamical friction ($1.8<z<2.6$) and then
merge with each other ($1.3<z<1.8$), leaving a compact disk ($z=1.3$).
Star formation rate is maximum ($\sim$ 30 $M_{\odot}$ ${\rm yr}^{-1}$)
at the epoch of the dissipative merging between two massive clumps
($z \sim 1.5$).
Gradual accretion of diffusely distributed gas onto the compact
disk results in the growth of a thin disk  between $z = 1.3$ and $z = 0.4$.
These results are basically in agreement with earlier numerical
results by Katz (1992) and Steinmetz \& M\"uller (1995).

Figure 2 shows how the metal-poor halo with [Fe/H]$\le-1.6$ is formed during
the collapse of the Galaxy. Clearly, a significant fraction of the metal-poor
stars ($\sim$ 50 \%) have already been formed within local small-scale density
maxima at early epoch $z>2.6$. These numerous clumps then merge with one
another and the debris stars constitute the outer part of the halo.
Later, the two massive subgalactic clumps are developed from the assembly of
smaller clumps and subsequently merge with each other. The stars
confined within the clumps are consequently disrupted and spread over the
inner region of the proto-Galaxy ($1.3<z<1.8$). As a consequence of this last
merging event, the flattened structure of the inner halo is formed.
The radial density distribution of the halo at $R \le 20$ kpc follows roughly
$\rho(R) \propto R^{-3.5}$, where $R$ is the distance from the center of the
disk (Bekki \& Chiba 1999). These structural properties of the simulated halo
are in good agreement with the corresponding observational results as reported
by CB.

Figure 3 shows that there is no significant correlation between [Fe/H] and $e$
for the stars with [Fe/H]$\le-0.6$, and that the existence of low eccentricity
($e<0.4$), low metallicity ([Fe/H]$<-1$) stars is successfully reproduced
in the present CDM model. This may be explained in a following manner.

First, as ELS argued, the rapid contraction of a gravitational potential within
a dynamical time ($\sim$ $10^{8}$ yr) results in the transformation of
initially nearly circular (smaller $e$) orbits to very eccentric (larger $e$)
ones. On the other hand, the eccentricities of the orbits remain basically
unchanged if the contraction is slow enough ($\sim$ $10^{9}$ yr). In the
present CDM model, the time scale for the contraction of the Galaxy
is lengthened by the expanding background universe
(the time scale for which a proto-Galactic
sphere with a turn-around radius of $\sim100$ kpc is of the order of Gyr),
 so that the process of star formation mainly
triggered by merging of small clumps is rather extended ($\sim$ 2 Gyr). Thus,
the orbital eccentricities of the metal-poor stars, once formed, are not
greatly influenced by the change of an overall gravitational potential of the
Galaxy.
Second, as we mentioned above, most of the metal-poor stars have been confined
within the massive clumps, where their orbits are gradually circularized due to
dissipative merging with smaller clumps and dynamical friction with the
dark halo particles. Thus, a finite fraction of the debris stars after the last
merging event preserve the orbital angular momentum of the clumps.
Both of these processes may give rise to the existence of low-$e$, low-[Fe/H]
stars in the simulated Galactic halo. 


To be more quantitative, we plot, in Figure 4, the cumulative $e$ distributions
of the metal-poor stars with [Fe/H]$\le-1.6$ (solid line) and
$-1.6<$[Fe/H]$\le-1.0$ (dotted line). For the halo component with
[Fe/H]$\le-1.6$, the fraction of the simulated low-$e$ stars with $e<0.4$ is
about 0.17, which is in good agreement with the observational result of about
0.2 (CY; CB). Also, as is consistent with the observational result, the
cumulative $e$ distribution in the intermediate abundance range
$-1.6<$[Fe/H]$\le-1.0$ is systematically larger than that for [Fe/H]$\le-1.6$.
The fraction of the simulated low-$e$ stars with $e<0.4$ in this abundance
range ($\sim 0.45$) is somewhat larger than the observation ($\sim 0.35$),
suggesting that the metal-weak thick disk component, which is emerged in this
intermediate abundance range, is somewhat over-produced. Besides this small
deviation from the observation, we conclude that the reported kinematic and
chemical properties of the Galactic halo are basically understandable in the
context of the CDM-based model for the Galaxy contraction.

\placefigure{fig-3}
\placefigure{fig-4}

\section{Discussion and conclusion}

Although both the ELS monolithic and SZ merger scenarios have offered the basic
ingredients for describing the early evolution of the Galaxy, either model
alone does not comprehensively explain the currently improved knowledge on
the fundamental properties of the halo (e.g., Freeman 1996; CB).
For example, the lack of the abundance gradient in the halo stars (Carney et
al. 1990; CY) and no significant correlation between [Fe/H] and $e$ (CY; CB)
are difficult to interpret in the context of the ELS scenario.
The SZ scenario seems unlikely to explain a large vertical gradient of
the mean rotational velocity $<V_{\phi}>$ in the halo component (CB).
It is also unclear how the rapidly rotating disk component subsequently formed
after the totally chaotic merging of ``Searle \& Zinn'' fragments.
In contrast, to explain the dual nature of the observed halo in its density,
kinematics, and age (Norris 1994; Carney et al. 1996; CB), one requires the
sort of hybrid picture, combining aspects of both the ELS and SZ scenarios.

As a possible candidate model to achieve the above hybrid picture, we have
considered the CDM model, which invokes both the hierarchical assembly of
subgalactic clumps and the dissipative process of gas inside the clumpy
protogalactic system.
As a first step toward understanding the formation of the Galactic halo
in this context, we have investigated the orbital properties of the stars
in the simulated Galactic halo, and have shown that the hierarchical merging
of CDM clumps in the course of the expansion and contraction of the overall
protogalactic sphere plays a vital role in determining the observed [Fe/H]-$e$
relation of the metal-poor stars. It is also found that the dissipative merging
of the clumps is important for the development of the characteristic structure
of the halo and also for the subsequently formed disk component (Bekki \& Chiba
2000).

While we have reproduced the most basic relation between metal abundances and
orbital eccentricities of the halo stars based on the currently favored picture
of galaxy formation, there are still a couple of points to be clarified for the
comprehensive understanding of the halo formation.
For example, Sommer-Larsen et al. (1997) reported that the velocity ellipsoid
of the metal-poor stars changes from radial anisotropy near the Sun to
tangential anisotropy in the outer part of the Galactic halo. This may be
explained via the anisotropic, {\it dissipative} merging between protogalactic
gas clouds in a collapsing galaxy (Theis 1997), but it is yet unsettled as to
whether the similar process is equally applied in the case of the CDM clumps
consisting of both dissipationless particles (dark matter and stars) and
dissipative gas.
Also, the non-kinematically selected sample of the nearby stars shows a
remarkable discontinuity of the mean rotational velocity, $V_{rot}$, at
[Fe/H]$\sim-1.7$: the stars at [Fe/H]$<-1.7$ shows an approximately constant
rotation, whereas those at [Fe/H]$>-1.7$ show the linear increase of $V_{rot}$
with increasing [Fe/H] (e.g., CB). We will further discuss in the forthcoming
paper  whether these other kinematic properties of the Galactic halo
are also explained by the dynamical evolution of the system of subgalactic
clumps seeded from the CDM density fluctuations.

\acknowledgments
We are grateful to  Edmund Bertschinger for allowing us to
use the COSMICS (Cosmological Initial Conditions and
Microwave Anisotropy Codes), which is a package
of fortran programs for generating Gaussian random initial
conditions for nonlinear structure formation simulations.

\clearpage


\figcaption{
Mass distribution of the forming Galaxy projected onto the $x-z$ plane at
each of redshifts  ($z = 25$, 2.6, 1.8, 1.3, 0.45, and 0)
in the present CDM model of galaxy formation.
Cyan and magenta colors represent gas and stars, respectively,
and dark matter particles are not plotted here for clarity. 
Each frame measures 126 kpc on a side.
\label{fig-1}}

\figcaption{
Mass distribution projected onto the $x-z$ plane
at each of redshifts  ($z = 2.6$, 1.8, 1.3, and 0)
for gaseous and stellar particles that finally become metal-poor stellar halo
component with metallicity [Fe/H]$\le-1.6$ at $z=0$.
Each frame measures 126 kpc on a side.
Here, if a progenitor of a metal-poor halo star (i.e., a progenitor of a
stellar particle with [Fe/H]$\le-1.6$ at $z = 0$) is still gaseous
at a given redshift, it is represented by cyan,
whereas if the progenitor already exists as a star at a given redshift,
it is represented by magenta. 
For example, about 50 (6) \% of the metal-poor halo stars are still gaseous
at $z$ = 2.6 (1.3).
Note that all particles at $z = 0$ are stars represented by magenta.
At $z=0$, the mass fraction of these metal-poor stars
distributed in the halo, compared with the total baryonic mass of the system,
is 0.014. The mean age of the stellar halo with  [Fe/H]$\le-1.6$
is about 10.5 Gyr.
\label{fig-2}}

\figcaption{
Relation between metal abundances ([Fe/H]) and orbital eccentricities ($e$)
for the simulated metal-poor stars with [Fe/H]$\le-0.6$.
Note the existence of low-$e$, low-[Fe/H] stars in the simulated halo.
\label{fig-3}}

\figcaption{
Cumulative $e$ distributions, $F(<e)$, in the two abundance
ranges of [Fe/H]$\le-1.6$ (solid line) 
and $-1.6<$[Fe/H]$\le-1.0$ (dotted one).
\label{fig-4}}


\clearpage
\begin{thebibliography}{}

\bibitem[]{} Beers,~T.~C., Chiba,~M., Yoshii,~Y., Platais,~I., Hanson,~R.~B.,
Fuchs,~B., \& Rossi,~S. 2000, \aj, submitted

\bibitem[]{} Bekki,~K. 1998, \apj, 502, L133

\bibitem[]{} Bekki,~K. \& Chiba,~M. 1999, in the proceedings of the MPA/ESO
meeting: the First Stars, ed. A.~Weiss (Berlin: Springer-Verlag), in press

\bibitem[]{} Bekki,~K. \& Chiba,~M. 2000, in preparation

\bibitem[Bekki \& Shioya 1998]{bs98} Bekki,~K., \& Shioya,~Y. 1998, \apj, 497,
108

\bibitem[]{} Bertschinger, E. 1995, astro-ph/9506070 


\bibitem[]{} Carney,~B.~W., Aguilar,~L, Latham,~D.~W., \& Laird,~J.~B.
1990, \aj, 99, 201


\bibitem[]{} Carney,~B.~W., Laird,~J.~B., Latham,~D.~W., \& Aguilar,~L.~A.
1996, \aj, 112, 668

\bibitem[]{} Chiba,~M., \& Yoshii,~Y. 1998, \aj, 115, 168 (CY)

\bibitem[]{} Chiba,~M., \& Beers,~T.~C. 2000, \aj, submitted (CB)

\bibitem[]{} Eggen,~O.~J., Lynden-Bell,~D., \& Sandage,~A.~R. 1962, \apj, 136,
 748 (ELS)

\bibitem[]{} Freeman,~K.~C. 1987, ARA\&A, 25, 603

\bibitem[]{} Freeman,~K.~C. 1996, in Formation of the Galactic Halo ....
Inside and Out, ASP Conference Series, eds. H.~Morrison \& A.~Sarajedini
(San Francisco: ASP), 3

\bibitem[]{} Helmi,~A., White,~S.~D.~M., de~Zeeuw,~P.~T., \& Zhao,~H.~S.
1999, \nat, 402, 53

\bibitem[]{} Katz,~N. 1992, \apj, 391, 502

\bibitem[]{} Katz,~N., \& Gunn,~J.~E. 1991, \apj, 377, 365

\bibitem[]{} Larson,~R.~B. 1974, \mnras. 166, 585

\bibitem[]{} Ma, C.-P., \& Bertschinger, E. 1995, \apj, 455, 7 

\bibitem[]{} Majewski,~S.~R. 1993, ARA\&A, 31, 575

\bibitem[]{} Norris,~J.~E. 1994, \apj, 431, 645

\bibitem[]{} Norris,~J.~E., Bessell,~M.~S., \& Pickles,~A.~J. 1985,
\apjs, 58, 463

\bibitem[Schmidt 1959]{sch} Schmidt,~M. 1959, \apj, 344, 685

\bibitem[Schwarz 1981]{sch81} Schwarz,~M.~P. 1981, \apj, 247, 77

\bibitem[]{} Searle,~L. \& Zinn,~R. 1978, \apj, 225, 357 (SZ)

\bibitem[]{} Sommer-Larsen,~J., Beers,~T.~C., Flynn,~C., Wilhelm,~R.,
\& Christensen,~P.~R., 1997 \apj, 481, 775


\bibitem[]{} Steinmetz,~M. \& M\"uller,~E. 1995, \mnras, 276, 549

\bibitem[]{} Steinmetz,~M., \& Navarro,~J.~F. 1999, \apj, 497, 108

\bibitem[Sugimoto et al. 1990]{sug90}
Sugimoto,~D., Chikada,~Y., Makino,~J., Ito,~T., Ebisuzaki,~T., \& 
Umemura, M. 1990, \nat, 345, 33

\bibitem[]{}
Theis,~C. 1997, in ASP Conf. Ser. 112, The History of the Milky Way
and Its Satellite System, ed. A.~Burkert, D.~H.~Hartmann, \& S.~A.~Majewski
(San Francisco: ASP), 35

\bibitem[]{} Urban,~S.~E., Corbin,~T.~E., Wycoff,~G.~L., Morton,~J.~C.,
Jackson,~E.~S., Zacharias,~M.~I., \& Hall,~D.~M. 1998, \aj, 115, 1212

\bibitem[]{} White,~S.~D.~M., \& Rees,~M.~J., 1978, \mnras, 183, 341


\bibitem[]{} Yoshii,~Y. \& Saio,~H. 1979, \pasj, 31, 339

\end{thebibliography}
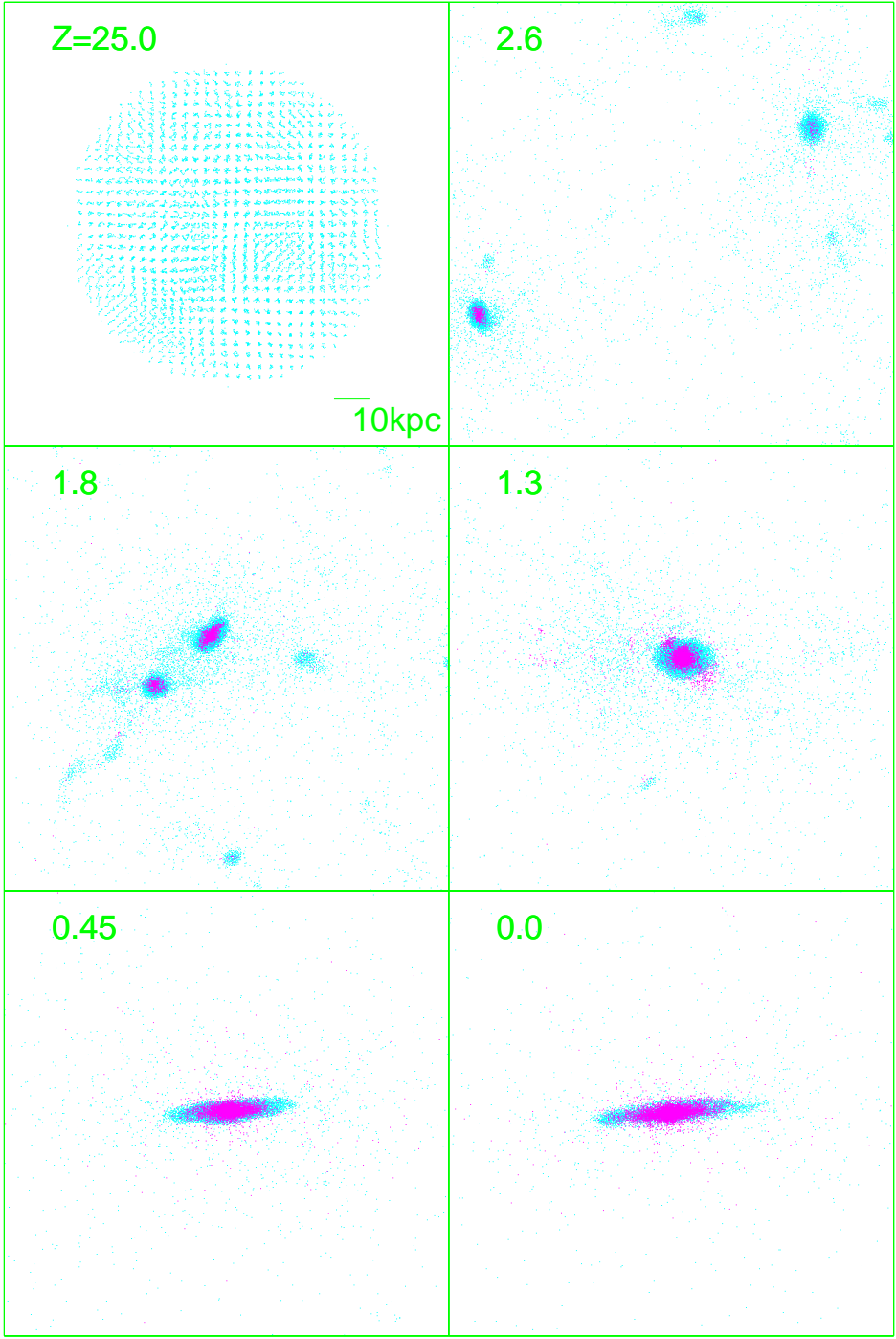
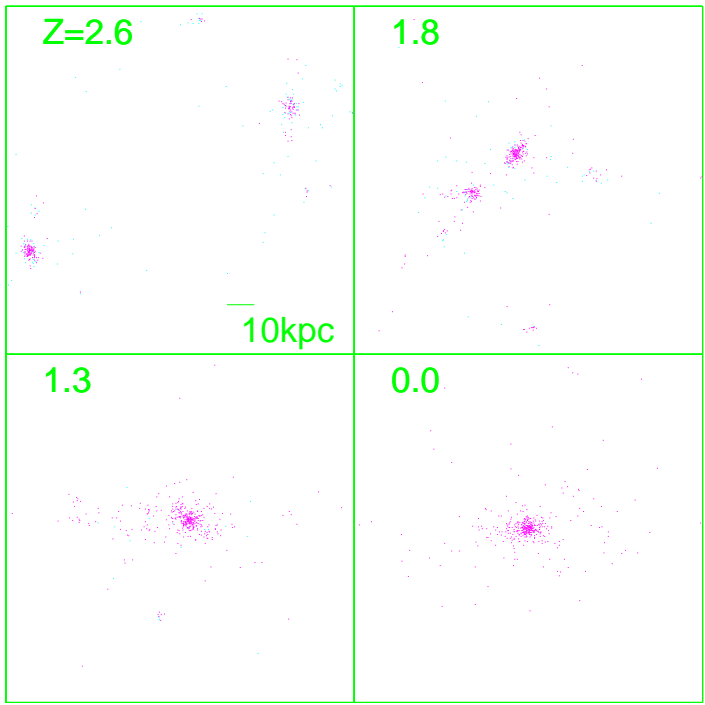
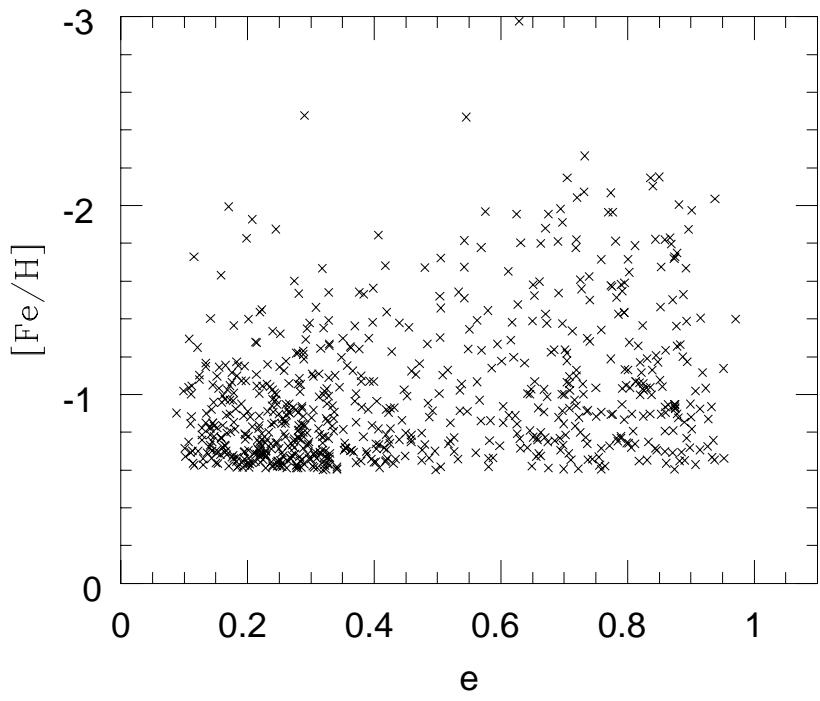
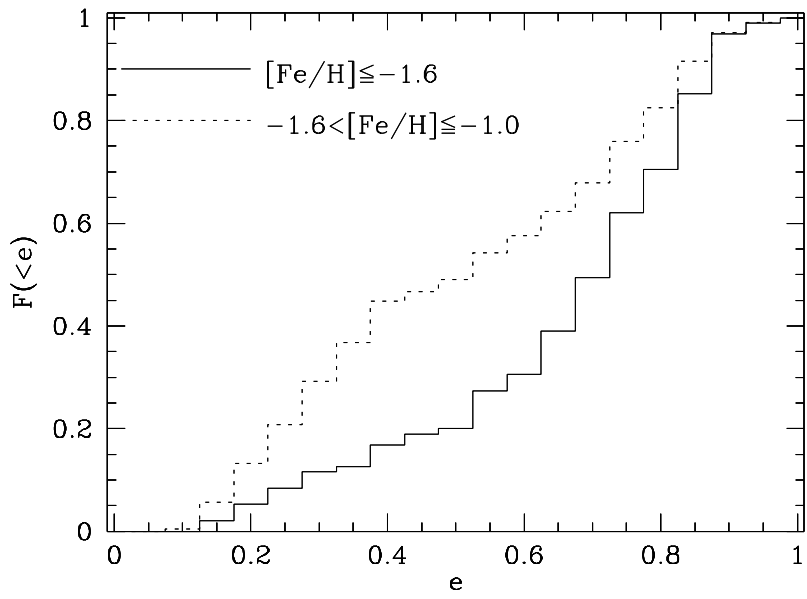
\end{document}